\documentclass[conference]{IEEEtran}

\usepackage{graphicx}
\usepackage{float}

\ifCLASSINFOpdf
\else
\fi
\usepackage[T1]{fontenc}

\usepackage[cmex10]{amsmath}

\begin{document}
%
\title{User Taglines: Alternative Presentations of Expertise and Interest in Social Media}

\author{\IEEEauthorblockN{Hemant Purohit}
\IEEEauthorblockA{Kno.e.sis Center, Computer Science and Engineering\\
Wright State University\\
Dayton, USA\\
hemant@knoesis.org}
\and
\IEEEauthorblockN{Alex Dow, Omar Alonso, Lei Duan, Kevin Haas}
\IEEEauthorblockA{Microsoft Silicon Valley\\
Microsoft Corp. \\
Mountain View, USA \\
\{first-name.last-name\}@microsoft.com}
}



\maketitle

\begin{abstract}
Web applications are increasingly showing recommended users from social media along with some descriptions, an attempt to show relevancy- \textit{why they are being shown}. For example, Twitter search for a topical keyword shows expert twitterers on the side for `whom to follow'. Google+ and Facebook also recommend users to follow or add to friend circle. Popular Internet newspaper- The Huffington Post shows Twitter experts on the side of an article for authoritative relevant tweets. The state of the art shows user profile bio as summary for Twitter experts, but it has issues with length constraints imposed by the user interface (UI) design, missing bio and sometimes funny profile bio. Alternatively, applications can use human generated user summary, but it will not scale. Therefore, \textit{we study the problem of automatic generation of informative expertise summary or taglines for Twitter experts in space constraint imposed by UI design}. We propose three methods for expertise summary generation: Occupation-Pattern based, Link-Triangulation based and User-Classification based, with the use of knowledge-enhanced computing approaches. We also propose methods for final summary selection for users with multiple candidates of generated summaries and evaluate results by user-study for both generation and selection tasks. The results of proposed tagline generation methods show 92.8\% good summaries with majority agreement in the best case and 70\% in the worst case while outperforming the state of the art up to 88\%. This study has implications in the area of expert profiling, user presentation and application design for engaging user experience.
\end{abstract}

\IEEEpeerreviewmaketitle

\section{Introduction}
\begin{figure}[h]
  \centerline{ \includegraphics[height=2.1in,width=3.0in]{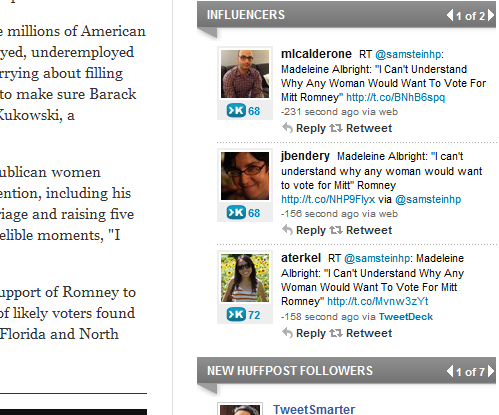} }
  \caption{ The Huffington Post showing Twitter experts list }
  \label{fig-huffpost-example}
\end{figure}

Recent growth in social media analytics has drawn interest from application designers to show Twitter experts or user recommendations with respect to target context. For example, a user searches for Madeleine Albright's comment on 2012 US Presidential election nominee Mitt Romney in a major search engine and she selects a document from the HuffingtonPost.com- a well-known Internet newspaper. Once reaching at The Huffington Post web page, she can see a number of related experts from Twitter on the right side (refer Figure \ref{fig-huffpost-example}). How can she understand- Who are these experts and why is she seeing a particular recommended user? There is an interaction gap unless the website shows summarization for these expert users, presenting \textit{`why a user is being shown'}. Similarly, as mentioned in the abstract, applications like Twitter search with `whom to follow' and Facebook's people search results or friend suggestions lack user summary for immediate interaction with this type of content on the website. Our focus here is expertise and interest presentation for Twitter experts in the form of short informative summarization, called as user tagline or summary. For example in the Figure \ref{fig-huffpost-example}, for first Twitter influencer `mlcalderone', a summary extracted from Twitter profile bio can be shown as \textit{`Senior Media Reporter for The Huffington Post'} as per our proposed approach.

The state of the art attempts to solve this problem by directly showing full bio of experts as a summary from the structured profile data. But, it leads to following issues-
\begin{enumerate}
\item  Space constraint on the UI design causing partial summary to be shown, 
\item  Empty summary field due to missing profile bio,
\item  Bio with no useful information, e.g., \textit{`Thanks for following me, guys!'}
\end{enumerate}
Alternatively, applications can use human resources to write informative expertise summaries but scalability will be an issue. Therefore, our goal is to design approach for automatically generating informative expertise summaries for Twitter experts in the space constraint imposed by UI.

In the earlier work on user summarization problem, researchers focused on user profiling from the viewpoint of topical personalization \cite{abel2011analyzing,hsu2012semantic,abel2011semantic}, and content recommendation \cite{deGemmis2008integrating,yarosh2012asking,zhou2012finding,reichling2009expert,chen2010short} applications. Analogous to user summarization, document summarization \cite{hahn2000the,madnani2007multiple,sharifi2010summarizing,daniel2009speech} was studied to summarize longer textual document. In both cases, structured and formal language content was available to understand context and generate computationally informative summaries without length constraints (e.g., topical tags for user modeling). Such summarization could be exploited for advanced analytics, such as content recommendation but may not be best suitable to present user expertise on UI, where human-friendly, readable and interesting summaries are needed.

We exploit \textit{`Meformer'} data (Self-descriptive, e.g., Twitter profile descriptions) as well as \textit{`Informer'} data (Others describe the target user, e.g., Wikipedia people pages) for approaching this problem. 
Traditional statistical summarization techniques have difficulty with the informal text of social media, therefore our algorithms make use of knowledge-bases and shallow Natural Language Processing (NLP).
Specifically, we use US Department of Labor Statistics reports and occupation titles collection \cite{miller1980work}, as well as Wikipedia knowledge-base for people's pages.
Using these knowledge-bases, we designed two candidate summary generation methods in addition to a default method for case of missing and noisy data nature:
\begin{enumerate}
\item	
\textbf{Occupation-Pattern based approach} first spots an occupation title (e.g., `author') collected from an occupation knowledge-base, followed by meaningful N-gram extraction that contain the title. This is faster and simpler method than completely depending on computationally expensive statistical learning of the language model. 
\item  \textbf{Link-Triangulation based approach} exploits Informer data from a knowledge-base (e.g., Wikipedia people pages), where this triangulation is based on three nodes: user's social network profile page (Twitter page), User's Wikipedia page and user's personal web page. It resolves user-identity in the knowledge-base by checking out-links to a user's personal web page from the remaining two nodes. After identity-resolution, we exploit structured metadata and content in the knowledge-base.
\item 
To overcome missing user profile information- bio or personal web page link, we propose a default \textbf{User-Classification based approach} to create user classes based on metrics of popularity, activity and content diffusion strength of a user in the network. 
\end{enumerate}

We note that some users may have multiple summary candidates, hence, we propose techniques for candidate summary selection based on three quality assessment principles: Readability, Interestingness and Specificity. We use a traditional linguistic approach for Readability measure \cite{kincaid1975derivation}, while Interestingness and Specificity are computationally modeled using modified tf-idf \cite{salton1988term} algorithm.

Our user-study based evaluation for generated summaries from the Occupation-Pattern based approach showed promising results of 74.5\% good summary with majority agreement for the best case, while our Link-Triangulation based approach showed 92.8\%, in comparison to 30\% for the baseline of user's full profile bio (refer Table \ref{table-evaluation-results-1}). We perform two phase experimentation where we compare baseline results in Phase-1 and test our methods on bigger datasets for generality in Phase-2. 

We list our contributions from this study here:
\begin{enumerate}
\item   First systematic approach to automatically generate expertise summary in space contraint of UI design
\item	Using knowledge-enhanced techniques for simpler and faster computation rather than sophisticated statistical language model
\item	Presenting significant improvement, up to 88\%, on the state of the art (based on statistics of good summaries with majority agreement for baseline as 30\% and for the best case of the proposed approaches as 91.6\% in Phase-1 experiments).
\end{enumerate}
The remaining paper describes related work in Section 2, problem formulation in Section 3, proposed methods in Section 4, corresponding experimentation and results in Section 5, followed by discussion on results in Section 6, future work in Section 7 and conclusion in Section 8.

\section{Related Work}
The problem of expertise presentation can be thought of as user summarization and presentation, which has been studied in various contexts of user modeling, mainly in user profiling, personalization, recommendation systems etc. Analogous to user summarization, document summarization also presents the document summary and has been focused study in the Information Retrieval research, which is also relevant literature to study in the current problem space.   

In the user profiling and personalization research, earlier work \cite{abel2011analyzing,hsu2012semantic,abel2011semantic} focused on user profiling by extracting topical tags from content for the purpose of creating computational user models rather than a presentation aspect of the user summarization. Notably, Abel et al. \cite{abel2011analyzing} proposed a framework for user modeling on Twitter which enriches the semantics of Twitter messages (tweets) and identifies topics and entities (e.g. person, events, products) mentioned in tweets and constructs hashtag-based, entity-based or topic-based user profiles. 
In the area of recommender systems, past studies \cite{deGemmis2008integrating,yarosh2012asking,zhou2012finding,reichling2009expert,chen2010short} were oriented for better approaches on identifying and ranking the experts for recommendation on contextual application of content- what features to target and how to capture domain specific requirements, such as news recommendation, product recommendation, advertisements etc. Gemmis et al. \cite{deGemmis2008integrating} proposed a semantic framework to understand user's content and created a content based user model to generate recommendations. 
Chen et al. \cite{chen2010short} studied content recommendation on Twitter to better direct user attention and therefore, explored three separate dimensions in designing such a recommender: content sources, topic interest models for users, and social voting.      

In the area of textual summarization, researchers studied various application specific summarization in the past literature \cite{hahn2000the,madnani2007multiple,sharifi2010summarizing,daniel2009speech}, mainly in the space of information retrieval for search, document summarization for storytelling in news, natural language processing etc. In such applications, the input data to summarize is long, structured and with formal English language; it allowed researchers to exploit NLP techniques and create sophisticated language models. The current problem space is challenged by informalism in user-generated content of social media, which limits the applicability of NLP techniques from past work.

In all of these relevant work, we observed contextual use-case of the user summarization whether by profiling for user model generations or for recommendation systems. Therefore, there is a clear need to address the problem space of automatic generation of expertise summary with a character limit imposed by UI design of applications.

\section{Problem Formulation}
We formulate the problem of expertise presentation for social media users in space limitation as a problem of generating informative textual summarization about a user in short description.

\textit{\textbf{Problem Statement}: Given a set of N experts E= \{$e_i$ | i= 1, 2, \dots, N\} in a social media community C of K (K > N) users, generate a short summary $d_i$ with maximum T characters for each of the expert $e_i$ in the set E.}

We use popular micro-blogging service Twitter in this study and limit to English text users, though proposed methods here can be extended for multi-lingual case which is out of scope and we plan it for future work.
We propose a solution to automatic expertise presentation problem by first generating candidate short summaries for an expert with length T characters, then selecting the suitable candidate for final summarization.

\section{Approach}
\label{Approach}
We describe data collection method, followed by candidate summary generation and final summary selection algorithms and lastly, the evaluation method.

\subsection{Data Collection}
We use two types of data in this study: 
\subsubsection{Meformer data}
In this data type, we consider data written by users themselves- user tweets as well as user profile metadata. Twitter Streaming API provides a random sample of the ongoing tweet stream, where each data point contains very rich metadata about tweet as well as the author. We store tweets and user metadata from the stream in a time slice for experimental study here. User metadata includes interests, location, number of tweets written, number of friends/followers etc. For sampling experts out of the user data set, we use a third party API service for expert-finding task, klout.com API here, but expert-finding task is not our primary focus, hence, any other mechanism can be used. We fetch expert scores for all users and rank the users, followed by extracting top k\% users for expertise summarization study, here k=30. 

\subsubsection{Informer data}
We consider information written by others for an expert. Therefore, we crawl knowledge-bases, such as Wikipedia and take its full data dump via API service. 

We also collect, occupation related lexicon using US Department of Labor Statistics reports and occupation titles collection \cite{miller1980work}, as described in detail in the following section for Occupation-Pattern based summary generation.

\subsection{Candidate Summary Generation}
We approach the summary generation task for expert users in our dataset by first investigating how well the Meformer data from user interest descriptions in the Twitter profiles can be used. We observed that on average 96\% of experts in our dataset had full bio. We noted that users write uninformative and funny bio which are not appropriate for expertise presentation, e.g.,

\textit{ `i been workin on the railroad,all the live long day... i been workin on the railroad jus to pass the time awaaaay...'. }

Also, the problem is further challenged when users write a long bio which exceeds the T characters threshold for summary length imposed by UI design. Thus, all these factors make it difficult to import full user bio directly. We propose following method to go beyond using the state of the art of using directly full user bio: 

\begin{figure}[h]
  \centerline{ \includegraphics[height=1.17in,width=3.5in]{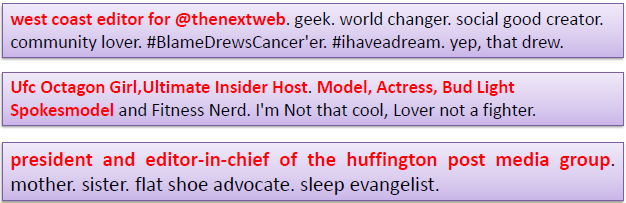} }
  \caption{ Examples of user profile bio with highlighted summary from Occupation-Pattern based approach }
  \label{fig-occupation-pattern-example}
\end{figure}

\subsubsection{Occupation-Pattern based Extraction}
We observed that user bio often have occupation titles, such as `author..', `editor-in-chief..' etc. Users write such occupation titles in a context which can be used to extract informative summary about the user, e.g., `editor of @TheNextWeb' (refer Figure \ref{fig-occupation-pattern-example} for examples). We take following steps in exploiting such patterns:

\begin{enumerate}
\item    \textit{Create occupation related lexicon}- By Collecting occupation titles using trusted knowledge-bases: US Department of Labor Statistics reports and occupation titles collection \cite{miller1980work} and Wikipedia's occupation categories. This lexicon is further augmented by human in the loop because of the informal nature of social media text, where users often write terms that are not understood by traditional English dictionary or sometimes, not considered formal occupation but they may be interesting to summarize user expertise, such as `footballer from Manchester United', `blogger at Huffingtonpost' etc.
\item	\textit{Filter the user bio data set}- By spotting occupation titles present in user bio, using the lexicon in step (1).
\item	\textit{Pre-process filtered user bio data set}- By removing noisy characters and words, such as multiple dots, new line characters, emails or contact information indicators- `contact us', `email us', `booking info' etc. and also by replacing URLs with proxy characters. 
\item    \textit{N-gram set creation from user bio}- Using Linguistic indicators of pause to tokenize user bio, here punctuations and conjunctions set: \{ , ; . / and \&\}.
\item    \textit{Extract meaningful N-gram set}- By selecting all the N-grams containing occupation patterns.
\item    \textit{Create a candidate summary}- By joining the members of the extracted N-gram set in step (5), create potential summaries with choice for final candidacy if the character length of summary does not exceed threshold T imposed by the UI design of the application.
\end{enumerate}
\textit{Intuition for joining the resultant N-grams for summary}: We follow orders of N-gram's positioning in the full bio because we trust the user's intelligence here to what is more important to describe him. For example,
\begin{itemize}
\item   User bio: \textit{`Tech journalist for All Things D. Oregonian transplanted to New York. Former BusinessWeek writer and columnist. Columbia grad.'}, and
\item   N-gram set: \textit{\{columnist, Tech journalist for All Things D, Former BusinessWeek writer\}}, then
\item   Candidate Summary/ Tagline: Starting with first potential candidate summary as \textit{`Tech journalist for All Things D, Former BusinessWeek writer, columnist'}, we note that it exceeds the T=70 character threshold in our experimentation, therefore, we create \textit{`Tech journalist for All Things D, Former BusinessWeek writer'} and \textit{`Former BusinessWeek writer, columnist'} candidate summaries for this user. 
\end{itemize}
We note that not all users have such occupation patterns in bio or sometimes bio is missing. Therefore, we exploit Informer data type in the next method.

\subsubsection{Link-Triangulation and Knowledge-base exploitation}
We note that expert users have the informative knowledge-base (e.g., Wikipedia) page because they are topical celebrities in a way and people have written informative content about them. But the challenge is to first resolve the user identity between an expert user in Twitter and the potential knowledge-base page. We exploit Informer data in two steps:
\begin{enumerate}
\item \textit{User Identity Resolution problem}- We use Link Triangulation approach (see Figure \ref{fig-wikipedia-triangulation-example}) to solve it, where the triangle is formed by three nodes: user's Twitter profile page, user's potential knowledge-base (Wikipedia) page/ content and user's personal web page. If personal page link in the user's Twitter profile points to an external link, which in turn, also an out-link in the knowledge-base content, then we conclude that the knowledge-base content belongs to the Twitter user. 
\begin{figure}[h]
  \centerline{ \includegraphics[height=1.3in,width=3.4in]{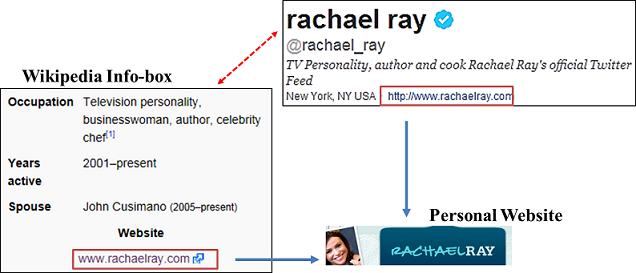} }
  \caption{ Link-Triangulation between the three nodes for a chef celebrity Rachael Ray: user profile on Twitter, Wikipedia page and user's personal page by exploiting the presence of a personal web page link. We use `Occupation' property of Wikipedia Infobox for summarization }
  \label{fig-wikipedia-triangulation-example}
\end{figure}
\item \textit{Meaningful summary extraction}- We exploit structured metadata content in the resolved knowledge-base content. We use Wikipedia Infobox, the informative summary box on the right side of a wiki page, and the `occupation' property metadata, for example `Actor, comedian, director, screenwriter'. We also extracted content from the first line of the Wiki page using the phrase after `is ', whenever Infobox property `occupation' was unavailable, but while parsing, we observed issue with random order of XML tags being used.
\end{enumerate}

At last, we apply length based normalization for the candidate summary as described in the step (6.) of the previous method by considering each occupation tag here as the N-gram.
We note that even after using both the aforementioned approaches we still remain with some experts without summarization, mainly due to unresolved user identity, missing structured metadata for `occupation' property, missing bio or uninformative user bio in the previous method etc. Therefore, we propose a default summary generation method in the next step.

\subsubsection{User Classification: Popularity, Diffusion Strength and Activity}
\begin{figure}[h]
  \centerline{ \includegraphics[height=1.92in,width=3.4in]{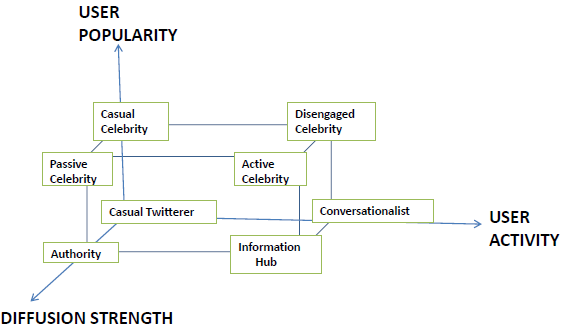} }
  \caption{ User Classification based on three principled metrics for default summary generation }
  \label{fig-user-classes}
\end{figure}

We use a Meformer data type in this approach, where we use user tweets. This approach makes use of tweets written by an expert as well as tweets in the interaction with this expert. We created three metrics on which users can be classified (see Figure \ref{fig-user-classes}): \\
- Popularity of the user to acknowledge its fame, \\
- Activity of the user in social media to consider a temporal aspect \\
- Diffusion Strength of user content to capture the ability to penetrate the user base of the social media communities.

We model these metrics in the current study for a user in a time slice as follows-
\begin{itemize}
\item \textit{Popularity} = Max normalized logarithmic value of the number of Twitter mentions of  the user
\item \textit{Activity} = Max normalized logarithmic value of the number of tweets written by the user
\item \textit{Diffusion Strength} = Max normalized logarithmic value of the number of retweeted tweets of the user
\end{itemize}
For a metric value V and maximum of metric values Max\_V, we compute normalized value Norm\_V as \\
  \centerline{  \textit { \textbf { Norm\_V = LOG(V+1)/LOG(Max\_V+1) } } \\ }

\begin{figure}[h]
  \centerline{ \includegraphics[height=1.9in,width=3.4in]{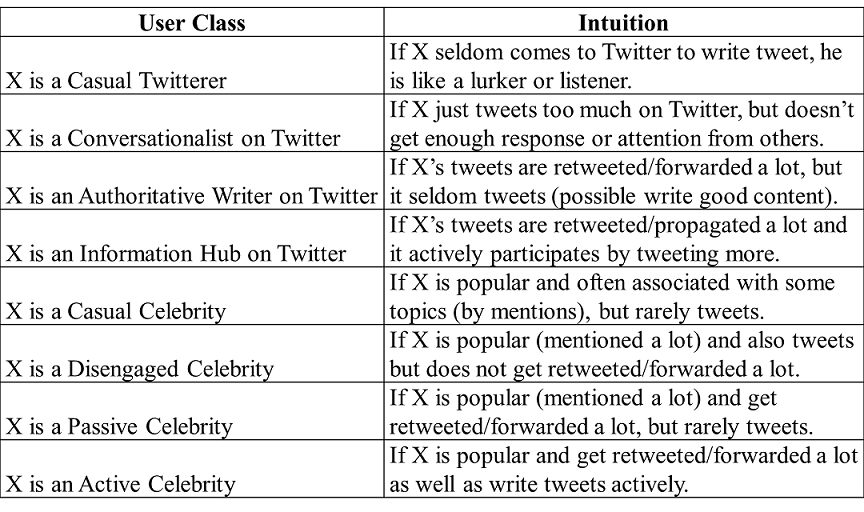} }
  \caption{ User Classification based expertise summary formats }
  \label{fig-user-classes-description}
\end{figure}

In the simplest classifier, we consider 50th percentile on each of the metrics to classify a user on two levels for each metric- low and high. In this way, we get 8 classes in the 3-dimensional space as shown in the Figure \ref{fig-user-classes} and the summary taglines for each of them is written as shown in the Figure \ref{fig-user-classes-description}.

Please note that this simple computation model of classifying users can be further advanced with more sophisticated metric computation. Our objective here is to design a philosophy of thought to present expertise rather than making really complex model. Such a simple model can be easily scalable.  This approach allows us to generate summary for all the users in generic classification. \\
As our next step, we propose algorithms to select the best candidate from output of first two summary generation methods and if none is available, we use the default method of user classification to generate the final summary.
\subsection{Final Summary Selection}
Choosing the best summary from a set of candidate summaries for a user can be a very subjective task. Also, the candidate summaries here are short in nature which presents another challenge to judge the quality, unlike document summarization problem where enough contexts is available. Therefore, we designed three principles to rely on for final summary selection being inspired from other problem spaces: 
\begin{itemize}
\item \textit{Readability}: how well the candidate can be read
\item \textit{Specificity}: what unique aspect is present in the candidate  
\item \textit{Interestingness}: how interesting is the candidate
\end{itemize}
We developed following methods to address computing model for each of these principles:

\subsubsection{Readability by Linguistic test}
We apply Flesch Reading Ease Scoring \cite{fleiss1971measuring} on each of the candidate summaries, where it assigns [0-100] score to a candidate summary, the higher the score, the better it is. This test computes the score based on syllables presence, complexity of words in the candidate etc. We observed that in many cases, it assigned 0 scores to candidates because of nature of fragmented sentence, presence of non-traditional English words resulting from social media conventions. 
Therefore, we plan to extend it further in the future work and skip the result statistics here.

\subsubsection{Specificity \& Interestingness by modified tf-idf approach}
We note that more specific information a summary contains, more likely it is to be informative and therefore, it is likely to generate more interest in the reader of the expertise presentation. Therefore, while reviewing analogous problem spaces, we observed that the Vector Space Model for document search in traditional Information Retrieval research computed a document's importance with respect to a corpus of documents and ranked relevant results to query term. In the similar way, a user can be thought of the query vector and candidate summaries of all users can be considered as a set of documents. 
Please note that the Vector Space model is not applied directly here because we are not formulating query vector of a user in the form of some user features 
and the set of relevant documents (summaries) to query (user) are known, as we know the candidate summaries belonging to a user. Therefore, our task is to find out the most important document (summary) by its own significance in the vector space of terms extracted from all the documents (summaries) of all the users as described below. We also normalize the summary scores by a maximum character limit imposed by UI design, in order to boost scores for candidates with an ideal length of utilizing available space when there is a comparison between the two candidates with marginal difference in scores. Final summary for a user is selected based on the highest score of the summary among candidates. The computation steps are described below:

Consider each candidate user summary as a document, say D and a unique term in that document, say t.

\begin{enumerate}
\item \textit{For term score:} \\
For each t in the document D with the total words as all\_words\_in\_D, compute the significance of  t, locally and globally:
\begin{itemize}
\item Locally, tf = term frequency in the form of frequency of t, say freq\_t\_in\_D 
\item Globally, idf = inverse-document frequency in the form of log ratio of total number of documents, say M and total number of documents containing t, say all\_D\_containing\_t: \\
\textit { \textbf {  tf-idf(t,D)=tf*idf=  }}  \\
\centerline{ \textit { \textbf { (freq\_t\_in\_D)*log(M/all\_D\_containing\_t) } } } %
\end{itemize}

\item	\textit{For document (summary) score:} \\
Aggregate the significance scores for each t, tf-idf(t,D) and normalize by document word length, all\_words\_in\_D. Aggregate function can be chosen in a more sophisticated way, but we chose it as SUM function: \\
\textit { \textbf {Score(D)= }} \\
\centerline{ \textit { \textbf { (AGGREGATE(tf-idf(t,D))/all\_words\_in\_D) } } }

\item	\textit{Normalized summary score by space constraint:} \\
Further normalize the score of D by the ratio of length of characters in D, say total\_characters\_in\_D and maximum character limit imposed by threshold from UI design, say T, to boost scores for the summary with length near to T (Please note that generated summaries have length <=T):  \\
 \textit { \textbf { Score'(D)=(Score(D)*total\_characters\_in\_D)/T) } } 
\end{enumerate}
We note that the space constraint based normalization for the summary scores had impact when there was a marginal difference in the scores otherwise it is not likely to affect an important high scoring short summary. 
\begin{figure}[h]
  \centerline{ \includegraphics[height=2.42in,width=3.32in]{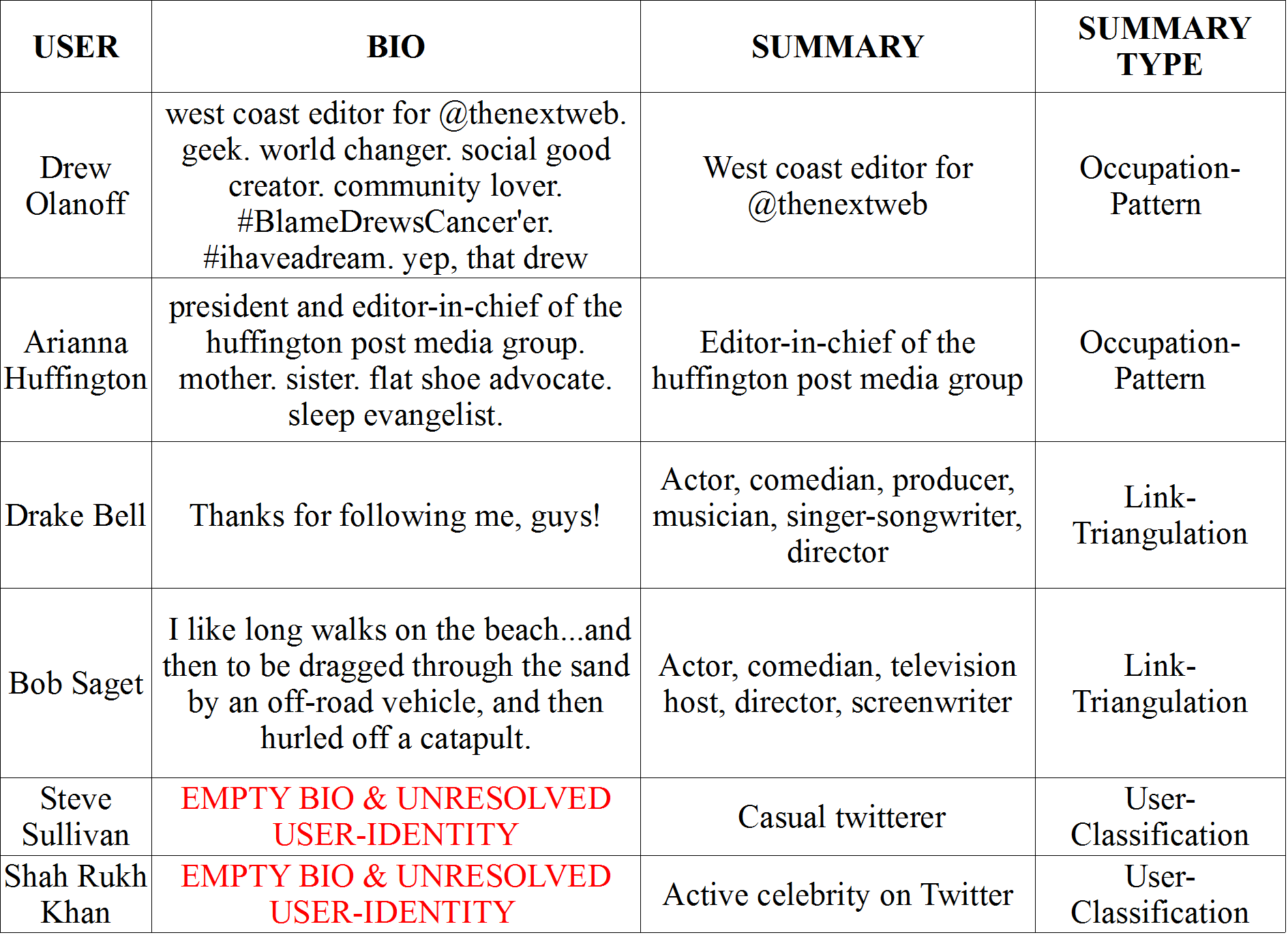} }
  \caption{ Examples of candidate expertise summary for various candidate generation methods }
  \label{fig-result-examples}
\end{figure}

\subsection{Evaluation}
After generating the candidate summaries and selecting the best one for a user automatically, we ask the question to evaluate quality of the generated summaries. As observed in the past literature, it is very difficult to evaluate the quality of natural language text without any labels in the data. In our study, we do not have any labeled data; therefore, we designed evaluation mechanism by asking a set of questions to human judges. 
We made sure that the judges were native English speakers for our evaluation tasks.

\subsubsection{User-Study}
In the candidate generation task, our objective is to evaluate if the generated summary is good to show expertise of a user. 
We show candidate summaries and the following set of questions to 3 judges for evaluating Occupation-Pattern based approach and setting up the baseline:
\begin{itemize}
\item \textit{ Q1. How good is the summary of expertise presentation? \\
Ans1. {2=very good, 1=good, 0=bad} } 
\item \textit{ Q2. How good the original bio is by itself? \\
Ans2. {2=very good, 1=good, 0=bad} } 
\item \textit{ Q3. How accurate is the original bio? \\
Ans3. {1=accurate, 0=misleading, -1=I don't know} } 
\end{itemize}
The purpose of the question Q1 is to judge the quality of automatically generated summary, while question Q2 is to test if we could directly use profile bio of the users, it helped us to set up a baseline. We use question Q3 to find out if users had misleading bio, for example, if a comedian writes \textit{`I'm president of US'}. We observed that there were not such cases during evaluation. For Link-Triangulation based approach, we ask the same question as Q1.

For candidate selection task, our objective is to evaluate if our approach can select a final summary in agreement with choice of human judges. We show various candidate summaries for a user to 5 judges and ask following question: 
\begin{itemize}
\item \textit{ Q1. Which one is the best summary to represent expertise description between two candidate summaries? \\
Ans. {1=candidate-1 is better, 2=candidate-2 is better, 0= both are almost same} }
\end{itemize}

\subsubsection{Reliable Metrics}
For reliability on the human judgment, we use majority inter-rater agreement for agreeable data samples from human and algorithmic judgements and Fleiss Kappa \cite{fleiss1971measuring}. Following we present statistics for evaluation:
\\
For candidate summary generation task-
\begin{itemize}
\item \textit{ Majority agreement for good summary samples \% } = Percentage of total samples, where at least two judges give scores as either 1 (good) or 2 (very good) for the question Q1
\item \textit{ Fleiss Kappa} = Another metric for inter-rater agreement
\item \textit{Good summaries \% } = Percentage of total judgments with scores as either of 1 (good) or 2 (very good) in the question Q1
\end{itemize}

For final summary selection task- 
\begin{itemize}
\item \textit{ Majority agreement for final summaries \% } =  Percentage of total samples, where at least 3 of 5 judges select the same candidate summary as by algorithmic selection
\item \textit{ Fleiss Kappa } = Another metric for inter-rater agreement
\item \textit{ Agreed final summaries \% } = Percentage of total judgments where one or more judges select the same final summary as by algorithmic selection
\end{itemize}

\section{Experiments}
We present experimental setup in this section for implementation and evaluation of the proposed methods for candidate summary generation and selection as well as the data set, baseline setup and results.
\subsection{Data set}
We took three snapshots of the Twitter data using its Streaming API for creating data sets, Set-S1, Set-S2 and Set-S3, as mentioned in the Table \ref{table-user-dataset-stats}. For each data set, we sampled expert users with the help of third party API service for expert-finding task, klout.com API here. We fetched expertise scores for all users, ranked them and extracted top 30\% for the expertise summarization study. We considered 30\% by the observation that application showing experts on UI is unlikely to reach more than 25\% of the users to show, therefore, 30\% is a safer threshold. 
We used a snapshot of Wikipedia's full data dump of the English database taken on May $1^{st}$, 2012. For occupation related lexicon creation, we took a snapshot of the occupation related census reports from the US Department of Labor Statistics and also, Wikipedia occupation category on June $15^{th}$, 2012 as described in the Data Collection section. Table \ref{table-user-dataset-stats} summarizes our data set for the experimentation.
\subsection{Experimentation}
We experimented for two fundamental tasks of our study- how to automatically generate good expertise summary? and how to select the best expertise summary if more than one candidate is available for an expert? We also set up the baseline as user's full profile bio as per the state of the art.\\
\textbf{\textit{ Candidate summary generation: }}
We took an iterative improvement approach on the proposed methods by establishing a baseline first using user profile bio and then, successive human evaluation on improved method versions for the automatic summarization. We developed our approaches using one data set Set-S1. We compare the results of the algorithmic output with baseline in Phase-1 on the data set Set-S1 in the Table \ref{table-evaluation-results-1} by user-study with 3 human judges. For testing efficacy of the proposed approaches in general, we show Phase-2 results where we experimented with a merged set of two different data sets, Set-S2 from a period earlier than Set-S1 and Set-S3 from a period after Set-S1, as shown in the Table \ref{table-user-dataset-stats}. \\
\textbf{\textit{ Final summary selection: }}
 Final summary for an expert was selected as the highest scoring candidate in the modified tf-idf approach described in the section \ref{Approach}. We evaluated efficacy of the selection task by showing two candidate summaries for each of the sample user to 5 judges and asking them to choose the best one. Then we compared the algorithmic results against human judgment. We show various result statistics in the Table \ref{table-evaluation-results-2} for this experimentation set. We note that there were really a less number of subjects to evaluate for the candidate selection task in our experiment because of less number of experts with multiple summaries in the data set during Phase-1. Also, we discarded the samples where users had given score as 0, implying `both the candidates are almost same for final summary' because it does not add any value to the evaluation question for choice. \\
\textbf{\textit{ Baseline: }}
 We consider full Twitter profile bio as the expert summary for baseline which is the state of the art. We evaluated the baseline using human evaluation method with the help of question Q2 as described in the Evaluation. We show baseline results in the Table \ref{table-evaluation-results-1}. We note that we could not find another type of baseline for the method of Link-Triangulation based approach and hence, we compare and contrast results directly with user profile bio based baseline. 
Given the subjectivity in the candidate selection task, we could not find a good baseline for evaluation of the selection algorithm and hence, we just report results of the user-study.

Tables \ref{table-evaluation-results-1} and \ref{table-evaluation-results-2} summarize various statistics from the user-study based evaluation for candidate summary generation methods as well as final summary selection.
\begin{table}[h]
\caption{ Statistics of expert users in three data sets }
  \centerline{ \includegraphics[height=0.8in,width=3.36in]{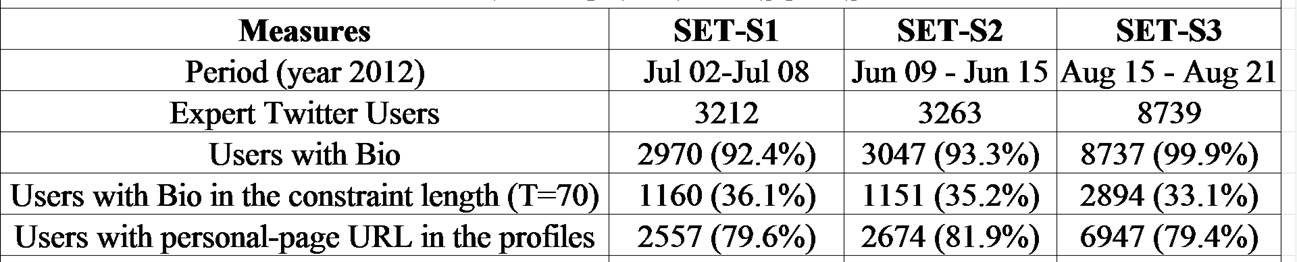} }
  \label{table-user-dataset-stats}
\end{table}

\section{Discussion}
We noted following observations from the Tables \ref{table-user-dataset-stats}, \ref{table-evaluation-results-1} and \ref{table-evaluation-results-2} for experimentation and results summary:
\begin{enumerate}

\item Table \ref{table-user-dataset-stats} shows the potential in exploiting profile bio and links based on average data availability of 96\% and 80\% respectively. Such higher availability may not be for a general Twitter users set, instead of experts. Also, not all the user bio are useful due to funny and uninformative bio data and also sometimes length (average is 100 characters in our data sets). For example, only average 35\% of bio qualified for T=70 characters limit in our experimentation (refer Table \ref{table-user-dataset-stats}).

\item     The baseline of full user profile bio shows 30\% good expertise summaries with majority agreement as shown in the Table \ref{table-evaluation-results-1}, thus, justifying the need for improvement in the state of the art in today's social media applications.

\item 	Phase-1 results in the Table \ref{table-evaluation-results-1} show that our candidate generation methods significantly outperform the baseline by producing 74.5\% good summaries with majority agreement for Occupation-Pattern based approach and 91.6\% for Link-Triangulation approach, in comparison to 30\% for the baseline. Also, Fleiss Kappa for the proposed methods is better than that for baseline. It may be due to irrelevant content alongside expertise information in the full bio, leading to confusion for the judges, as observed from some of the judges' comments, e.g., \textit{`Im a MAVIN, ARTIST, SONY/ATV Songwriter and a child of GOD. 323 Entertainment/MAVIN Records. PR Rep: @Wunmie09 Bookings: @TeeBillz323 323musicent@gmail.com'}.
\begin{table}[h]
\caption{ A user-study based evaluation results for candidate expertise summary generation for question Q1 }
  \centerline{ \includegraphics[height=1.6in,width=3.3in]{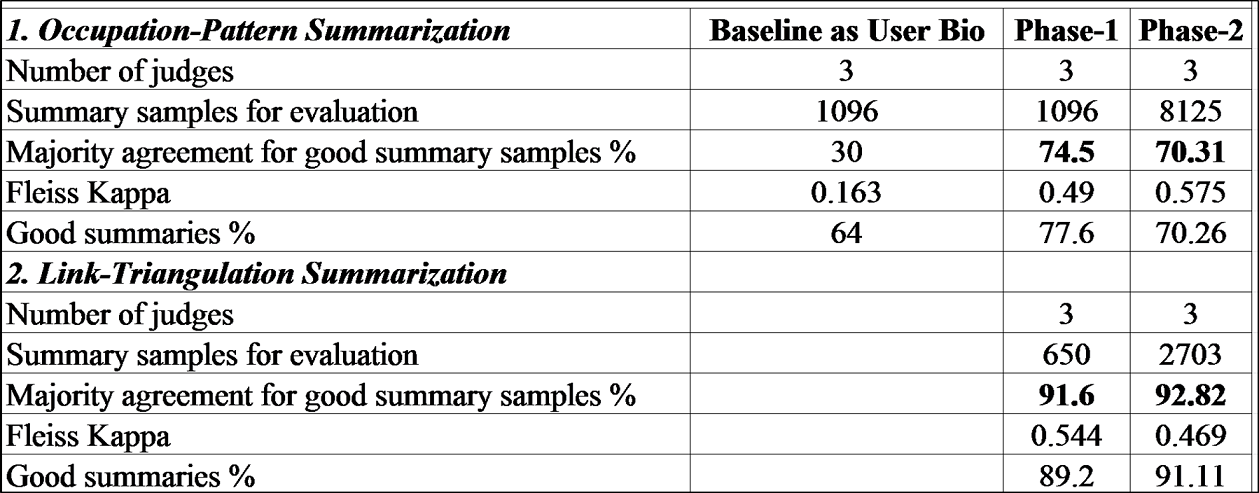} }
  \label{table-evaluation-results-1}
\end{table}
\begin{table}[h]
\caption{ A user-study based evaluation results for final expertise summary selection tasks for question Q1 }
  \centerline{ \includegraphics[height=0.7in,width=2.9in]{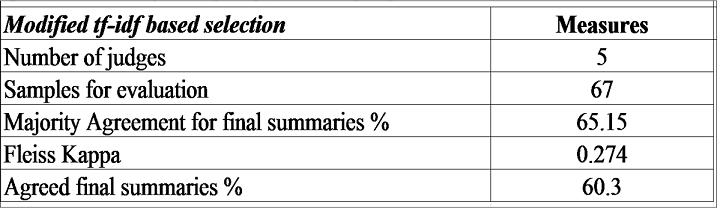} }
  \label{table-evaluation-results-2}
\end{table}
\item 	Phase-2 focused on generic evaluation of the proposed candidate generation approaches by evaluating results on the larger dataset. It showed 70.3\% good summaries with majority agreement for Occupation-Pattern based and 92.82\% for Link-Triangulation based approaches.

\item 	Even after applying Occupation-Pattern and Link-Triangulation based methods, there are lots of experts without summaries due to missing or uninformative profile bio or unresolved user identity. Even when the user identity resolution is done, sometimes the informative data in the knowledge-base is missing or ill-formatted to extract meaningful summary. Therefore, our method to generate default tagline by user classification is as important as the other two. In the nutshell, all the three methods can play important complementary role. The Figure \ref{fig-result-examples} shows some examples of the expertise summaries for all the methods.

\item 	Evaluation of the candidate selection approach in the Table \ref{table-evaluation-results-2} shows 65.1\% majority agreement between the automatic selection of final summary candidate and by human judgment. We note that the samples for the evaluation are lesser in this case due to availability of lesser user samples with multiple candidate summaries. We also note that the best summary selection from a set of given candidate summaries is a very subjective task and therefore, future work can focus on it.

\item 	\textbf{Applications}: Our study can be applied in various domains for user presentation and expert profiling, such as search, recommendation etc. Also, in another application such as coordination during disaster situations where meaningful user taglines from our approach can help tech savvy emergency responder team to quickly engage with Twitter community users who want to help, based on their expertise. 

\end{enumerate}

\section{Future Work}
We plan to explore a semantic approach in creating expert summary, focussing on semantic association of the summary parts with respect to domain specific application need. 
Work on creation of news titles and snippets in the Search results are also useful, which capture interesting and important part of documents in small space. We plan to extend readability measure of user summaries based on language modeling on the web corpus.
Future studies can also extend proposed methods for multi-lingual expertise summaries.

\section{Conclusion}
This paper presents the first systematic study of expertise presentation in constrained space for social media applications by automatically generating candidate summaries and selecting final summary/ tagline. We proposed various methods to generate candidate summaries for an expert, followed by principles and methods to select the best candidate. We also described user-study based evaluation of the proposed approaches. This study not only contributes to being first to systematically outline the methods to approach this problem, but also presents significant results, outperforming the state of the art. This study will have implications for expertise presentation in search, recommendation as well as expert profiling and application design.

\section*{Acknowledgment}

We thank reviewers for their valuable comments and colleagues for encouraging feedback, especially Dr. Shubha Nabar, Vasilis Kandylas, Prof. Amit Sheth and Wenbo Wang.



%

\bibliographystyle{IEEEtran}
\bibliographystyle{alpha}
\bibliography{myReferences}

\end{document}